\documentstyle[11pt,newpasp,twoside,epsf]{article}
\def\edcomment#1{\iffalse\marginpar{\raggedright\sl#1\/}\else\relax\fi}
\reversemarginpar

\begin{document}
\title{New Era in Likelihood Analyses of the Local Group Acceleration}
\author{Micha{\l} Chodorowski \& Pawe{\l} Cieciel\c{a}g}
\affil{Copernicus Astronomical Center, Bartycka 18, 00-716 Warsaw, Poland}

\begin{abstract} 
In maximum-likelihood analyses of the Local Group (LG) acceleration,
the object describing nonlinear effects is the {\em coherence
function} (CF), i.e.\ the cross-correlation coefficient of the Fourier
modes of the velocity and gravity fields. We study the CF numerically,
using a hydrodynamic code. The only cosmological parameter that the
CF is strongly sensitive to is the normalization $\sigma_8$ of the
underlying density field. We provide an analytical fit for the CF as a
function of $\sigma_8$ and the wavevector.  The characteristic
decoherence scale which our formula predicts is an order of magnitude
smaller than that determined by Strauss et al. This implies that
present likelihood constraints on cosmological parameters from
analyses of the LG acceleration are significantly tighter than
hitherto reported.
\end{abstract}

\section{Introduction}
\label{sec:intro}
Comparisons between the CMB dipole and the Local Group (LG)
gravitational acceleration can serve not only as a test for the
kinematic origin of the former but also as a constraint on
cosmological parameters. A commonly applied method of constraining the
parameters by the LG velocity--gravity comparison is a
maximum-likelihood analysis (Strauss et al. 1992, hereafter S92). In
such an analysis, a proper object describing nonlinear effects is the
{\em coherence function} (CF),
\begin{equation} 
C({\bf k}) = \frac{\langle {\bf g}_{\bf k} \cdot {\bf v}_{\bf k}^\star
\rangle}{\langle |{\bf g}_{\bf k}|^2 \rangle^{1/2} \langle |{\bf
v}_{\bf k}|^2 \rangle^{1/2}} \,,
\label{eq:dec_def} 
\end{equation}
where ${\bf g}_{\bf k}$ and ${\bf v}_{\bf k}$ are the Fourier
components of the gravity and velocity fields, and $\langle \ldots
\rangle$ means the ensemble averaging. 

\section{Results of numerical simulations}
\label{sec:res}
We model cold dark matter as a pressureless fluid. We chose to use a
grid-based code rather than a $N$-body code because it directly
produces a volume-weighted velocity field, as required here.  

We checked that the CF depends on $\Omega_m$ very weakly. Next, we
tested its dependence on the normalization of the power spectrum. We
found that the dependence of the function on $\sigma_8$ and the
wavevector can be well modelled as
\begin{equation}
C(k) = \exp{(-ak)} \,, 
\label{eq:func_fit}
\end{equation}
where 
\begin{equation}
a = \left\{ \begin{array}{ll}
0.757~\sigma_8^2 & \mbox{for $\sigma_8 \le 0.3$} \,, \\
-0.059 + 0.423~\sigma_8 & \mbox{for $0.3 < \sigma_8 \le 1.0$} \,.
\end{array} \right. 
\label{eq:a_fit}
\end{equation}
In the left panel of Figure~1 we show the CF for the values of
$\sigma_8$ given by the cluster normalization, for flat models with
$\Omega_m = 1$ and $\Omega_m = 0.3$, respectively. Dotted lines show
the function from numerical simulations, while solid ones are our
fits. The fits are good; we  checked that for other values of
$\sigma_8$ they are as good as those shown here.

The CF has been modelled by S92, who calibrated it so as to fit the
results of N-body simulations of a standard CDM cosmology. In the
right panel of Figure~1 we show S92's prediction for the function, as
well as our results, for the standard CDM power spectrum and
$\sigma_8$ normalization of S92 ($0.625$). The discrepancy of our
results with the formula of S92 is drastic! Instead of a
characteristic decoherence scale of $4.5$ $h^{-1}\,\hbox{Mpc}$ (S92),
our formula suggests a fraction of a megaparsec.

Greater coherence of the LG velocity with the LG gravity implies
tighter constraints on cosmological parameters that can be obtained
from their comparison. Thus, {\sl in likelihood analyses of the LG
acceleration, the value of $\beta$ can be determined with
significantly greater precision than is currently believed}.

\begin{figure}
\plottwo{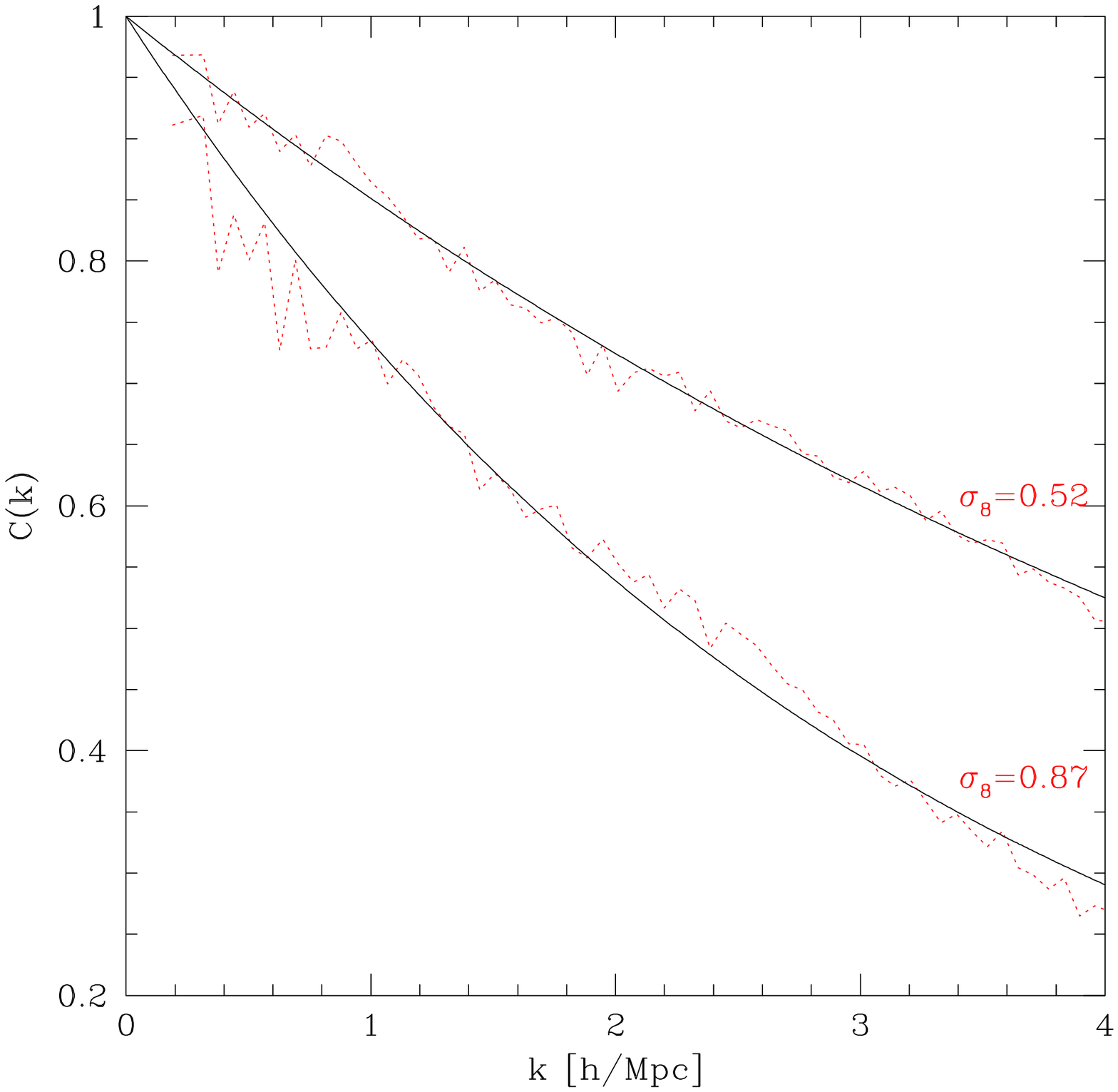}{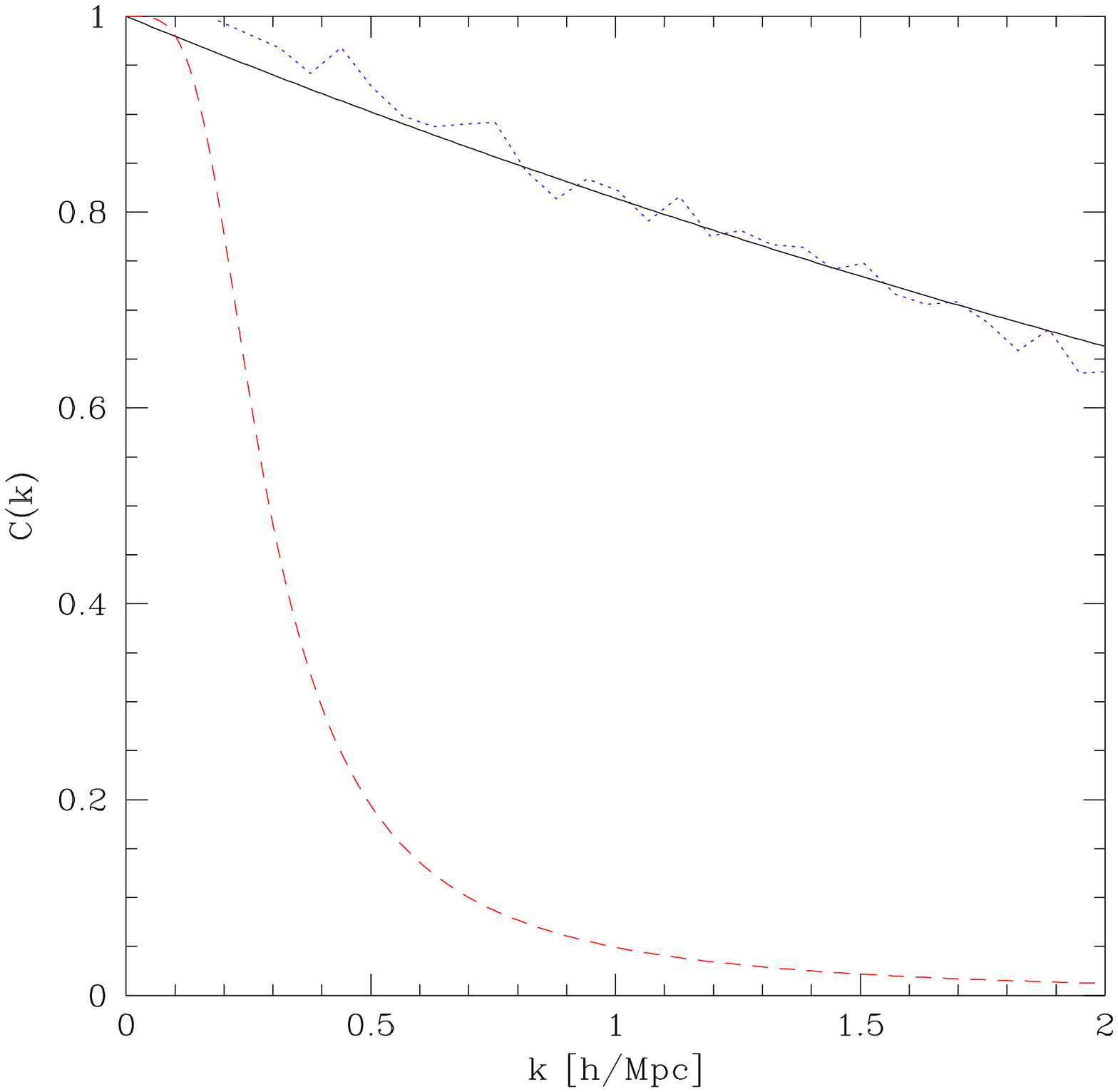}
\caption{
\emph{Left:\/} The coherence function for cluster normalizations of
the PSCz power spectrum. Dotted lines show the function from numerical
simulations, while solid ones are our fits. \emph{Right:\/} The
function for a standard CDM cosmological model with $\sigma_8 =
0.625$. Dotted: numerical simulations, solid: our fit. The dashed
line is the formula~(18) of S92, with $r_c = 4.5$
$h^{-1}\,\hbox{Mpc}$.}
\end{figure}

\end{document}